\documentclass[aps,prd,fleqn,superscriptaddress]{revtex4}
\usepackage{graphicx,color,natbib}
\usepackage{amsmath,amssymb,amsfonts}

\usepackage{epsfig,epsf}
\usepackage{dcolumn}
\usepackage{float}
\usepackage{bm}

\newcommand{\bse}{\begin{subequations}}
\newcommand{\ese}{\end{subequations}}
\newcommand{\be}{\begin{equation}}
\newcommand{\ee}{\end{equation}}
\newcommand{\bea}{\begin{eqnarray}}
\newcommand{\eea}{\end{eqnarray}}
\newcommand{\ba}{\begin{array}}
\newcommand{\ea}{\end{array}}

\input amssym.def
\input amssym.tex

\usepackage[colorlinks=true, linkcolor=blue, bookmarks=true]{hyperref}

\begin{document}

\title{Resource and stability near a critical point from the quantum information perspective}

\author{Mohammad Ali-Akbari\footnote{m\_aliakbari@sbu.ac.ir}}
\affiliation{Department of Physics, Shahid Beheshti University, 1983969411, Tehran, Iran
}
\author{Mahsa Lezgi\footnote{s\_lezgi@sbu.ac.ir}}
\affiliation{Department of Physics, Shahid Beheshti University, 1983969411, Tehran, Iran
}

\begin{abstract}
In the presence of chemical potential and temperature, we holographically study subregion complexity in a non-conformal quantum field theory with a critical point. We propose a new interpretation according to which the states, needing (more) less information to be specified, characterize the (un) stable thermodynamical solutions. We observe the increasing and decreasing effects of chemical potential and temperature on holographic subregion complexity, respectively. These two opposite behaviors lead to a point where subregion complexity of the mixed state is 
the same as this value for a zero temperature conformal field theory. We also present a new description of the difference between the minimum and the maximum value (the value near the critical point) of holographic subregion complexity as a resource for doing computational work to prepare the state near the critical point from the state far from it. We also calculate the critical exponent.
\end{abstract}
\maketitle
\tableofcontents

\section{Introduction}

The critical endpoint is a favorite subject for investigation and speculation. Near the critical point, different first and second derivatives of the free energy go to zero or diverge which is known as power laws. Describing the critical exponent associated with these power laws is one of the most important areas of critical phenomena. A number of  models have been introduced to analyze the critical point and its properties in strongly coupled field theories. However, for example, lattice calculations are difficult in the presence of finite chemical potential \cite{lattice}.   

Gauge-gravity duality is a useful tool for the study of strongly coupling systems according to which a strongly coupled gauge theory in $d$-dimensional space-time corresponds to a classical gravity in $d+1$-dimensional space-time \cite{CasalderreySolana:2011us}. Different phenomena in strongly-coupled field theories from condensed matter physics to low-energy quantum chromodynamics (QCD), are described by this strong-weak duality \cite{erdmenger}.
 
What has attracted a lot of attention in the literature is connections between quantum information and quantum gravity in the gauge/gravity duality framework. This interesting area started with the Hubney-Ryu-Takayanagi proposal for entanglement entropy, as a measure of quantum correlation of a pure quantum state. A simple geometrical prescription has been introduced to describe entanglement entropy and its properties and it passes many tests successfully \cite{Takayanagi,Nishioka:2009un}. 

Quantum complexity is another main concept in information theory. It refers to the time and space resources needed to do a computation efficiently \cite{mi}. In other words, the quantum complexity is the minimum number of elementary operations, quantum gates (unitary operators in the context of quantum field theory), required to produce a state of interest from a fixed reference state \cite{mi2}.

In the holographic view, it has been proposed that the quantum complexity of the boundary field theory state has a geometric description in the gravitational dual background in association with the interior geometry of the black hole \cite{cc}. Two different recipes have been introduced for describing the complexity of pure states holographically, based on characterizing the size of the black hole interior with its spatial volume (CV) and its action (CA) \cite{cv,ca}. The CV proposal states that the complexity is the volume of the extremal/maximal volume of a codimensional-one hypersurface $\mathcal{B}$ in the bulk ending on a time slice of the boundary 
\begin{equation}
{\cal{C}}=\frac{V(\mathcal{B})}{\hat{L} G_N},
\label{cv}
\end{equation}
where $\hat{L}$ is a length scale associated with the geometry and $G_{N}$ is the Newton's constant. Inspired by the Hubney-Ryu-Takayanagi proposal, this conjecture is generalized for mixed states \cite{alishahiha}. For a subsystem $A$ on the boundary, the complexity equals to the volume of codimension-one hypersurface enclosed by the Hubney-Ryu-Takayanagi surface, $\gamma_{A}$, 
\begin{equation}
{\cal{C}}_A=\frac{V_{\gamma_A}}{8\pi R G_N},
\label{sc}
\end{equation}
where $\cal{C}_{A}$ is called holographic subregion complexity (HSC) and $R$ is $AdS_5$ radius. Some works on HSC for various gravity models can be found in \cite{co1,co2,co3,co4,co5,co6,co7,co8}.

Description of complexity as the entropy of an auxiliary classical system, where unitary operators evolve, opened a new window to a topic known as thermodynamics of complexity \cite{sus11}. It was known that the lack of entropy or {\it{negentropy}}, is a resource for doing a work. In parallel, a definition for complexity was defined. This definition, as a resource in computational work, is called {\it{uncomplexity}} and refers to the difference between maximum possible complexity and the actual complexity. In this case, work refers to doing directed computation, a computation with a goal which is done by a computational machine \cite{sus11, sus1}. 

In this paper we would like to study the behavior of HSC near the critical point. Hence, we consider a field theory with a critical point which is characterized by temperature and chemical potential. Its phase diagram has a first order phase transition line that ends in a critical point. Its gravity dual is a charged black hole which is discussed in section \ref{21}. We calculate HSC in this model, in section \ref{211} and present our numerical results in section \ref{2111}.

\section{Review on the background}\label{21}

We are interested in investigation of the critical phenomena of a strongly coupled plasma using the gauge/gravity framework. Hence, we review a holographic geometry  dual to a four-dimensional field theory with critical point. This metric is supported by a five-dimensional action of the 1RCBH model \cite{gub}. The solution can be summarized as
\bea
\label{metric}
ds^2 &=& e^{2 A(r)} (-h(r) dt^2+d{\vec{x}}^2) + \frac{e^{2 B(r)}}{h(r)} dr^2,\nonumber\\
A(r) &=& \text{ln} r + \frac{1}{6} \text{ln} (1+\frac{Q^2}{r^2}) ,\cr
B(r) &=& - \text{ln} r - \frac{1}{3} \text{ln} (1+\frac{Q^2}{r^2}) ,\cr 
h(r) &=& 1-\frac{M^2}{r^2(r^2+Q^2)} ,\cr 
\phi(r) &=& - \sqrt{\frac{2}{3}} \text{ln} (1+\frac{Q^2}{r^2}), \nonumber\\
A_t(r) &=& \bigg{(} - \frac{M Q}{r^2+Q^2}+\frac{M Q}{r_H^2+Q^2}\bigg{)}, 
\eea
where $A_{t}$ is the time component of the gauge field and $\phi$ is dilaton field. $M$ and $Q$ are the black hole mass and charge. AdS radius is set to one throughout the paper. The boundary is located at $r \rightarrow \infty$ and $r_{H}$ is the horizon and is obtained as follows
\be
\label{horizon}
r_H=\sqrt{\frac{\sqrt{Q^4+4 M^2}-Q^2}{2}} .
\ee 
The temperature in the dual field theory is 
\begin{equation}
T=\frac{\sqrt{-{g_{tt}}' {g^{rr}}'}}{4 \pi}\bigg{|}_{r=r_H} = \frac{Q^2+2 r_H^2}{2 \pi \sqrt{Q^2 + r_H^2}}.
\label{tem}
\end{equation}
Due to the gauge field in the bulk, the chemical potential in the field theory is
\begin{equation}
\mu=\lim_{r \to \infty} A_t = \frac{Q r_H}{\sqrt{Q^2 + r_H^2}}.
\label{mu}
\end{equation}
The relation between parameters in the bulk and the boundary field theory, using \eqref{tem} and \eqref{mu} is obtained
\be
\frac{Q}{r_H}=\sqrt{2} \bigg{(} \frac{1\pm \sqrt{1-(\frac{\sqrt{2}}{\pi}\frac{\mu}{T})^2}}{\frac{\sqrt{2}}{\pi}\frac{\mu}{T}} \bigg{)}.
\ee
\label{re}
Therefore, the background contains two different branches of variables $\frac{Q}{r_{H}}$ for each value of $\frac{\mu}{T}$ which specify stable and unstable branches of solutions. This is a sign of the existence of a first order phase transition in field theory. The two branches merge at the critical point where $\left(\frac{\mu}{T}\right)_{c}=\frac{\pi}{\sqrt{2}}$ $\left(\frac{Q}{r_H}=\sqrt{2}\right)$ and the branch with the parameters satisfying $\frac{Q}{r_{H}}<\sqrt{2}$ is thermodynamically stable \cite{hajar}.

This background has been studied in different papers. For example as the holographic dual of QCD in \cite{DeWolfe} and by computing the bulk viscosity and baryon conductivity, the critical exponent was obtained \cite{DeWolfe2}. The value of the critical exponent was confirmed in \cite{quasi} and \cite{bahman} by investigation of quasi-normal modes and entanglement of purification, respectively. Moreover, in \cite{hajar} the behavior of complexity using CV and CA conjectures has been studied near the critical point and it has been found that the critical exponent can be achieved by using various time scales up to very small error. 

\section{Holographic subregion complexity}\label{211}
\begin{figure}[H]
\centering
\includegraphics[width=65 mm]{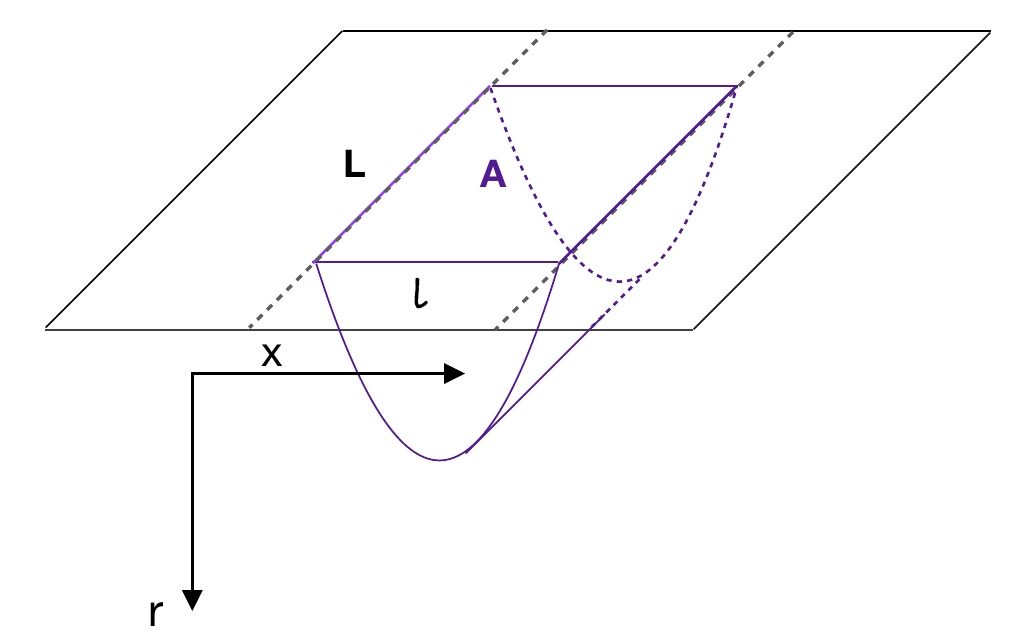} 
\caption{A strip entangling surface of length $l$ and width $L\rightarrow \infty$.}
\label{0}
\end{figure}

The CV conjecture extends to complexity for mixed states which is known as HSC \cite{alishahiha}. Motivated by the Hubney-Ryu-Takayanagi proposal, HSC states that the complexity of the mixed state defined as a region $A$ on the boundary is dual to the volume enclosed by the minimal hyper-surface $\gamma_{A}$, appearing in the calculation of holographic entanglement entropy, i.e. equation \eqref{sc}. We consider a rectangular region on the boundary which has width $L\rightarrow \infty$, see Fig \ref{0}. This subsystem is defined by $-\frac{l}{2}<x_{1}(\equiv x)<\frac{l}{2}$ and $-\frac{L}{2}<x_{2},x_{3}<\frac{L}{2}$ at a given time. Using \eqref{metric}, we easily find the area of the minimal surface as follows
\begin{equation}
S=\frac{L^2}{4G_{5}}\int_{-\frac{l}{2}}^{\frac{l}{2}}e^{3A(r)}\sqrt{1+\frac{e^{2(B(r)-A(r))}}{h(r)}r'(x)^2}dx,
\label{action}
\end{equation}
where $r(x)$ (equivalently $x(r)$) is the profile of the minimal surface which is obtained by using the constant of motion
\begin{equation}
x(r)=\int_{r_{*}}^{r}\frac{e^{B(r)-A(r)}}{\sqrt{h(r)(e^{6(A(r)-A(r_{*}))}-1)}}dr,
\label{profile}
\end{equation}
where $r_{*}$ is the turning point of the minimal surface and lies at $x=0$. By integrating the inside of the minimal surface using slicing the bulk with planes of constant $r$, the volume enclosed by $\gamma_{A}$ is found
\begin{equation}
V(r_{*})=2L^{2}\int_{r_{*}}^{\infty}\frac{e^{3A(r)+B(r)}}{\sqrt{h(r)}}x(r)dr.
\label{volume}
\end{equation}
The volume is divergent and we need to introduce a normalized version of volume using \eqref{sc} as follows 
\begin{equation}
C\equiv\frac{8\pi R G_{N}(\mathcal{C}-\mathcal{C}_{AdS})}{L^{2}}=\frac{V-V_{AdS}}{L^{2}},
\label{volume2}
\end{equation}
where $\mathcal{C}$ and $\mathcal{C}_{AdS}$ are the HSC for $A$ in \eqref{metric} and AdS geometry respectively. The volumes are for the same boundary region such that $V$ in \eqref{volume} reduces to $V_{AdS}$ by setting $M$ and $Q$ equal to zero. We use the above definition to discuss our finding.

\section{Numerical results}\label{2111}

\begin{figure}[H]
\centering
\includegraphics[width=65 mm]{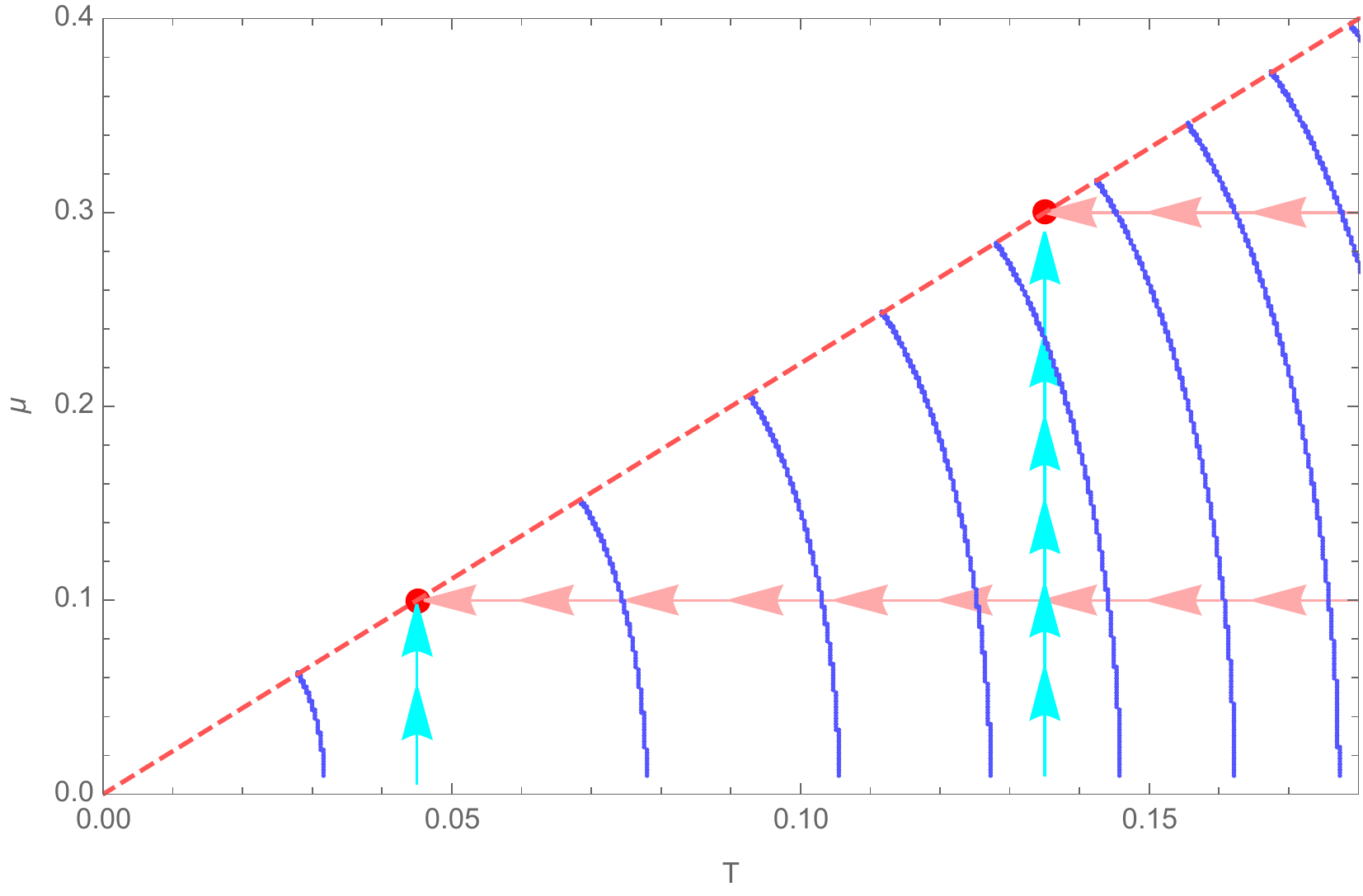} 
\caption{Chemical potential as a function of temperature. The red dashed line is the critical line, $\mu=\frac{\pi}{\sqrt{2}}T$. The blue curves are thermodynamically stable data of $\mu$ and $T$. The light blue vertical lines and the pink horizontal lines indicate moving towards critical points in the stable region of data, with constant $T$ and $\mu$, respectively.}
\label{00}
\end{figure}

The field theory we have considered is characterized by temperature $T$ and chemical potential $\mu$. Since the underlying theory is a conformal field theory, the observables of field theory we consider are characterized by the ration of $\frac{\mu}{T}$, as expected. For instance, similar to this theory but in another holographic model, we have made analytical calculations in \cite{co5} and obtained a rescale HSC in terms of dimensionless quantities of theory. Here, we will present our results from the numerical calculation of normalized HSC, which is introduced in \eqref{volume2}. Hence, we do not have explicit dependence of $\frac{\mu}{T}$ and we will study normalized HSC as a function of $T$, $\mu$ and $\frac{\mu}{T}$. In Fig \ref{00}, we have plotted $\mu$ as a function of  $T$. There is a region of thermodynamically stable solutions, under the critical line i.e. $\mu=\frac{\pi}{\sqrt{2}}T$, the red dashed line. We can move toward critical points in directions with constant $T$ (light blue lines) or constant $\mu$ (pink lines), or in directions with arbitrary values of $T$ and $\mu$ in the allowed stable region of data (blue curves). In the upcoming discussion, we study the behavior of $C$ as a function of $T$, $\mu$ and $\frac{\mu}{T}$ and approach the critical point in different ways, in fixed $T$ or $\mu$ directions and directions in which, with varying values of $\frac{\mu}{T}$, we move towards $\left(\frac{\mu}{T}\right)_{c}$. In other words, although it is clear that the observables of the system are functions of $\frac{\mu}{T}$, we consider the special cases of constant $T$ and $\mu$, to complete the discussion.
We expect that the critical exponent as a property of the theory does not depend on how the theory is explored. 

\begin{figure}[H]
\centering
\includegraphics[width=57 mm]{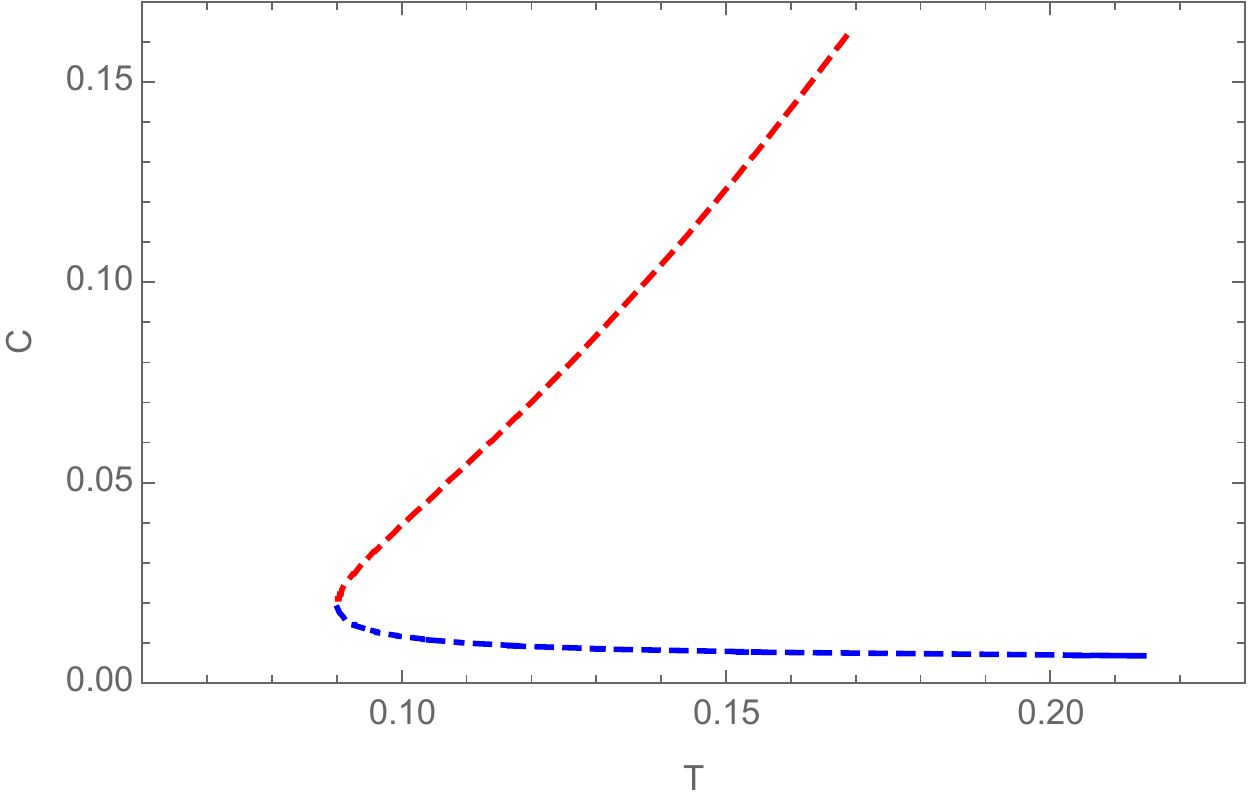} 
\includegraphics[width=57 mm]{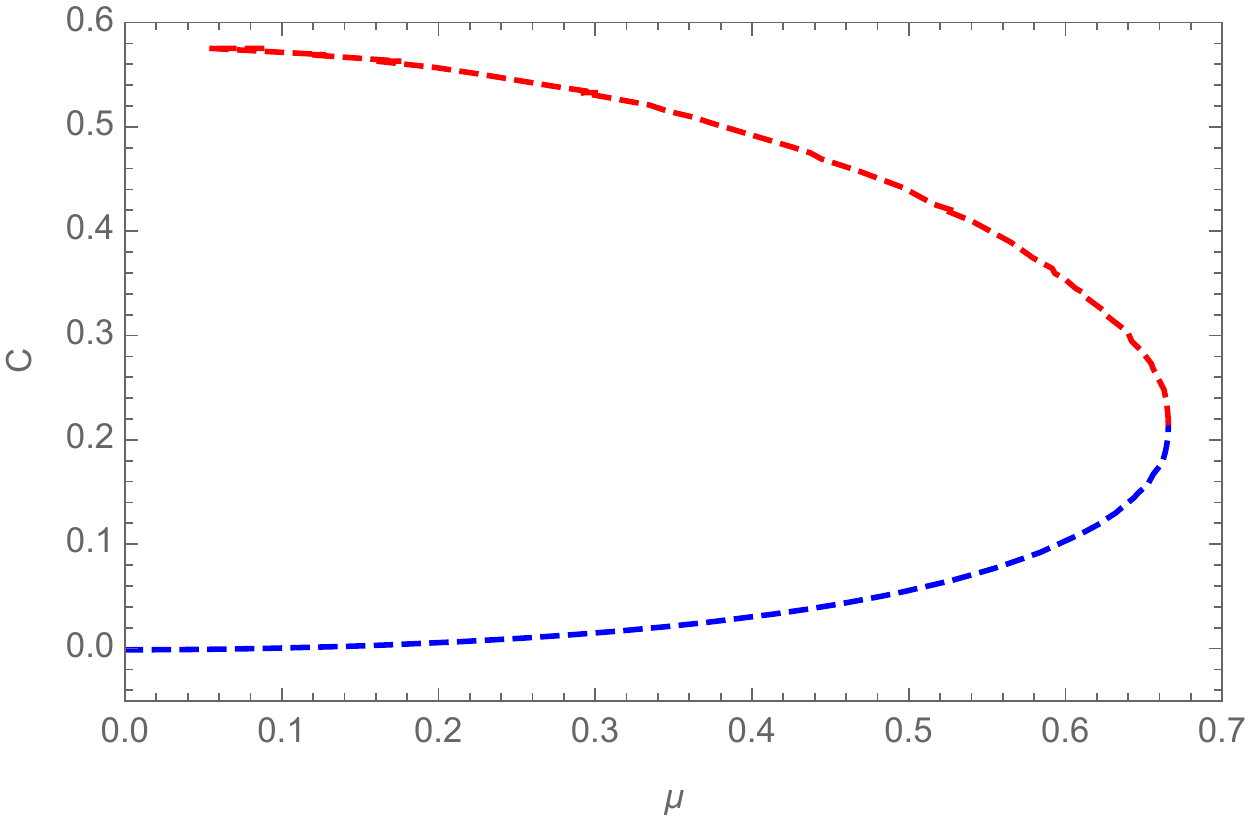}
\includegraphics[width=57 mm]{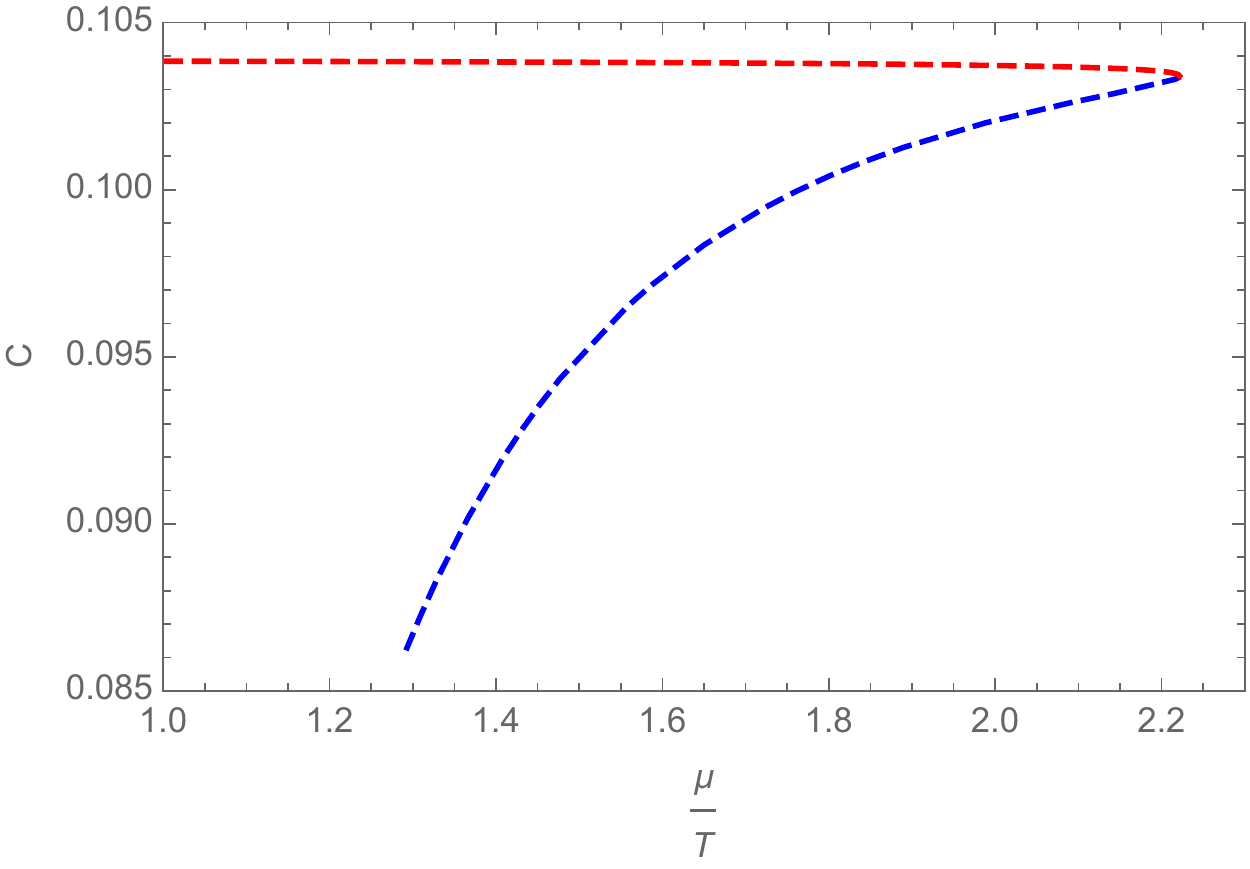}
\caption{Left: $C$ as a function of $T$ at $\mu=0.2$ and $l=0.1$ for stable (blue) and unstable (red) branches. Middle: $C$ as a function of $\mu$ at $T=0.3$ and $l=0.1$ for stable (blue) and unstable (red) branches. Right: $C$ in terms of $\frac{\mu}{T}$ for $l=0.1$ for stable (blue) and unstable (red) branches.}
\label{1}
\end{figure}

At first, we would like to investigate whether the thermodynamically stable and unstable solutions can be specified from a quantum information point of view. For this purpose, we study the behavior of $C$ as a function of $T$, $\mu$ and $\frac{\mu}{T}$. In Fig \ref{1}, we have plotted $C$ in terms of $T$ for a fixed value of $\mu$ (left panel), in terms of $\mu$ for a fixed value of $T$ (middle panel) and in terms of $\frac{\mu}{T}$ (right panel). It shows that $C$ is a double-value function, the  thermodynamically stable branch (blue curve) and thermodynamically unstable solution (red curve), as expected. The two branches intersect at the critical point where $T$ and $\mu$ satisfies $\left(\frac{\mu}{T}\right)_{c}=\frac{\pi}{\sqrt{2}}$ . It is interesting that between two branches with the same $T$ (left), $\mu$ (middle) and $\frac{\mu}{T}$ (right), the solution with a smaller value of $C$ belongs to the thermodynamically stable solution. This means, according to \eqref{volume2}, the amount of information it takes to determine the considered state is less. In other words, we want to conclude that the state specified with less information, characterizes the stable branch. This result is consistent with the results reported in  an anisotropic holographic model studied in \cite{co8}. 

\begin{figure}[H]
\centering
\includegraphics[width=67 mm]{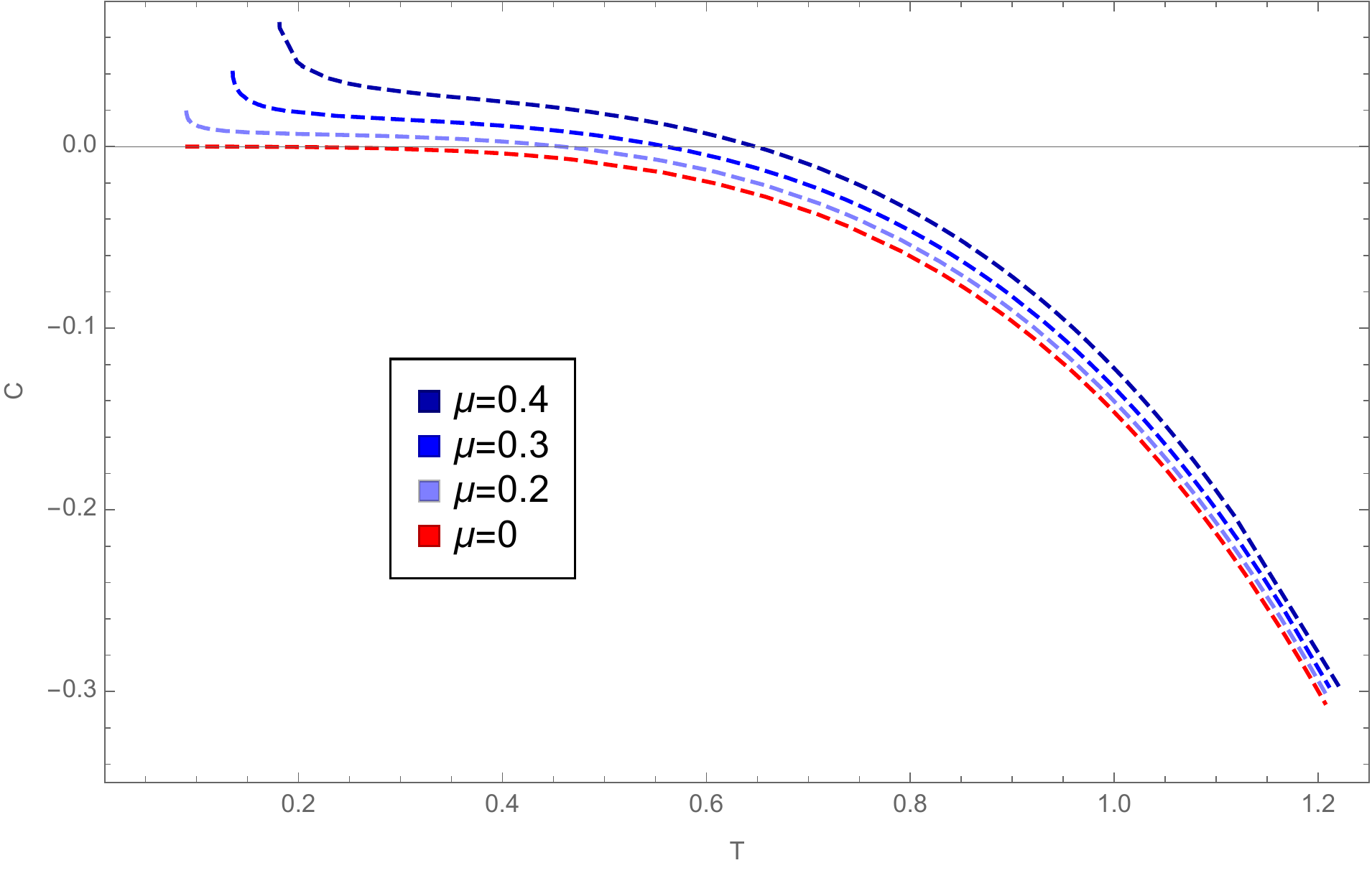} 
\includegraphics[width=67 mm]{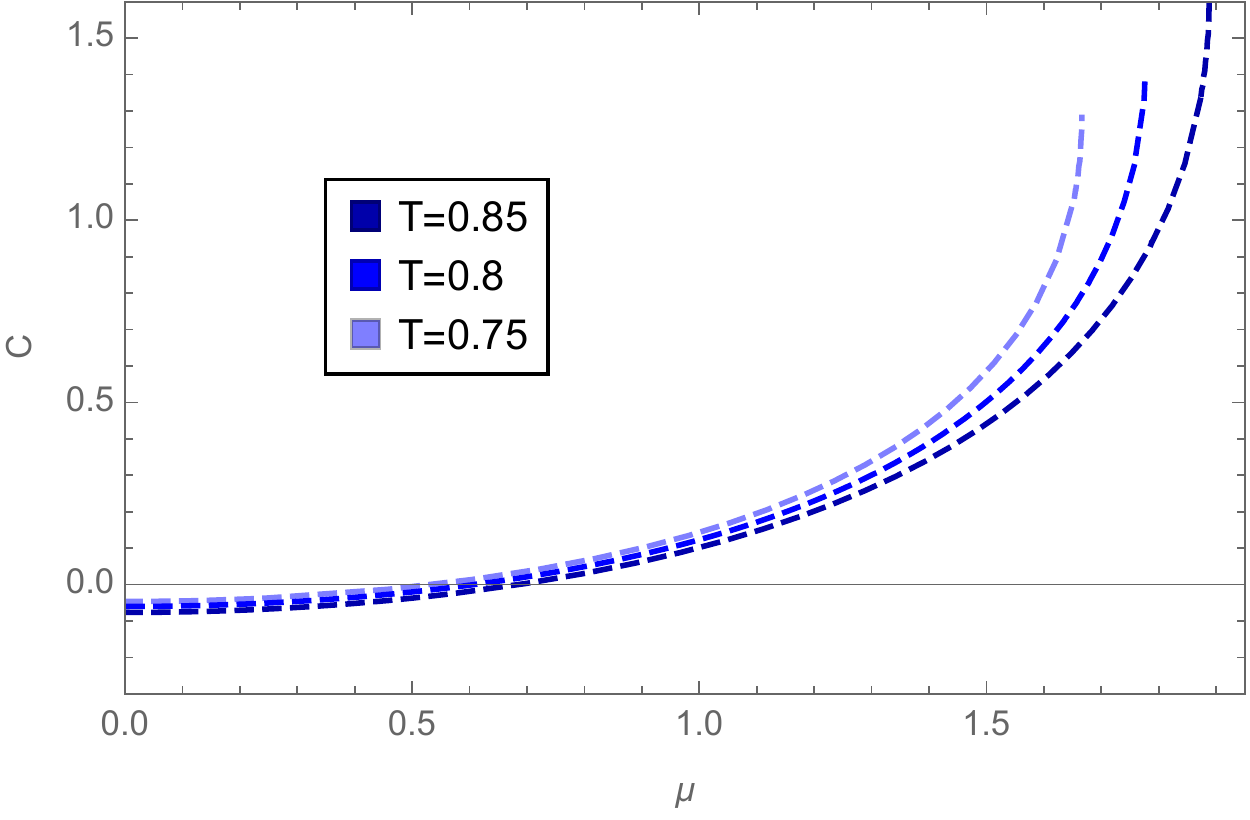} 
\caption{Left: $C$ as a function of temperature at $l=0.1$ for stable branches in different values of $\mu$. The red curve is the behavior of $C$ for AdS black hole metric. Right: $C$ in terms of chemical potential at $l=0.1$ for stable branches in different values of $T$.}
\label{2}
\end{figure}

In Fig \ref{2}, we have focused on thermodynamically stable (or stable from the point of view of HSC) branches. In the left panel, we have plotted $C$ as a function of $T$ for three fixed values of $\mu$ (blue curves) and for a conformal field theory which is dual to AdS black hole solution and it is obtained by setting $Q=0$ in \eqref{volume2} (red curve). There exists two regions, one where $T$ is dominant in high temperature limit and one for which $\mu$ is dominant, in low temperatures. In high temperatures, $C$ for different values of $\mu$ tends to $C$ for the conformal thermal field theory (red curve). In fact, for large enough values of $T$, $C$ is independent of $\mu$. In low temperatures, the difference between curves for various $\mu$ becomes manifest. Also in this region, with a decrease in $T$ and as we approach the critical points for different values of $\mu$, $C$ has a power-law behavior, as expected. In the following, we will study this behavior and obtain the critical exponent.

An interesting observation is the change of sign of $C$. In order to explain it, first note that with the increase of $T$, $C$ decreases. Here, our previous argument, in \cite{co7,co8}, about this behavior with varying temperature based on an ensemble of microstates corresponding to a given mixed macrostate is valid. At zero temperature there is one microstate, which is called {\it{unique configuration}}, and with increasing temperature, the number of microstates corresponding to the mixed macrostate increases. Therefore, the state is specified with less information than is done at zero temperature. On the other hand, with increasing $\mu$, $C$ increases. In low temperatures, $C$ is positive which means according to \eqref{volume2}, $\mathcal{C}>\mathcal{C}_{AdS}$. In this region, the effect of $\mu$ is significant. We expect that HSC of a zero temperature mixed state should be more than that of the thermal mixed state. However, the effect of $\mu$ in increasing the HSC of our mixed state prevails over the effect of $T$ and therefore $C$ is positive. But with the increase of $T$, its effect on the behavior of $C$ increases and we can see in the large enough values of $T$, the sign of $C$ changes and, $\mathcal{C}<\mathcal{C}_{AdS}$. It means the HSC of the zero temperature mixed state becomes more than the HSC of the thermal mixed state.
Interestingly, there exists a point where $\mathcal{C}=\mathcal{C}_{AdS}$ which means the decreasing effect of $T$ on HSC compensates for the increasing effect of $\mu$ on it. Therefore, at this point despite having $T$ and $\mu$, the HSC of the two states is the same. We do not expect a state where $\mathcal{C}=\mathcal{C}_{AdS}$, in the presence of $T$ and $\mu$. However, due to the opposite effect of $T$ and $\mu$ on HSC, there is a state in which these two effects cancel each other out. Similar results are obtained from the right panel of Fig \ref{2}, in which $C$ has been plotted in terms of $\mu$ for three  
different values of $T$.

\begin{figure}[H]
\centering
\includegraphics[width=55 mm]{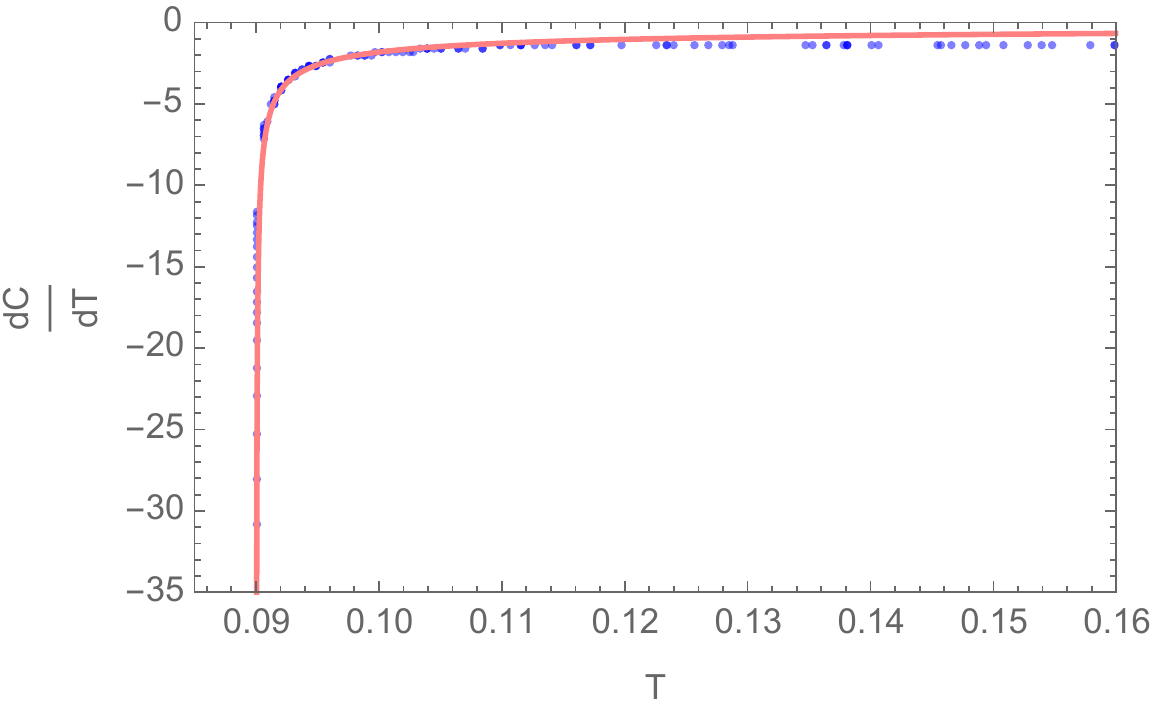} 
\includegraphics[width=55 mm]{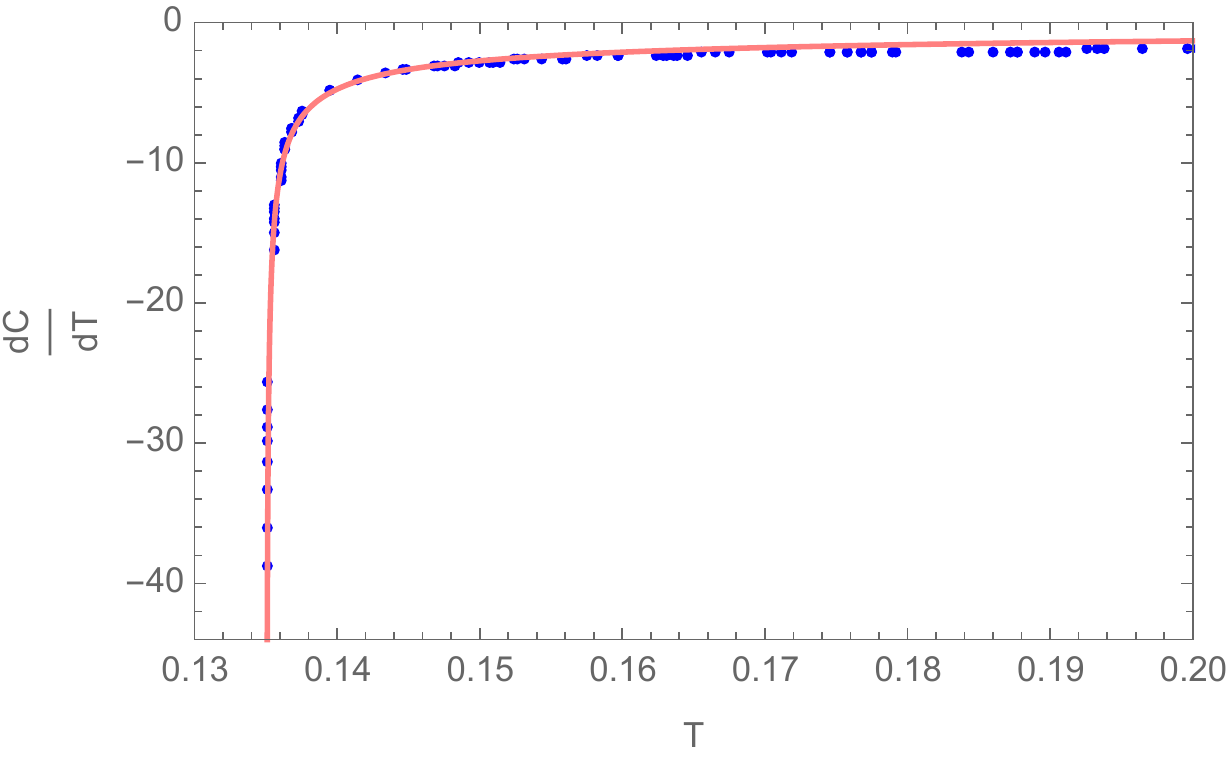} 
\includegraphics[width=55 mm]{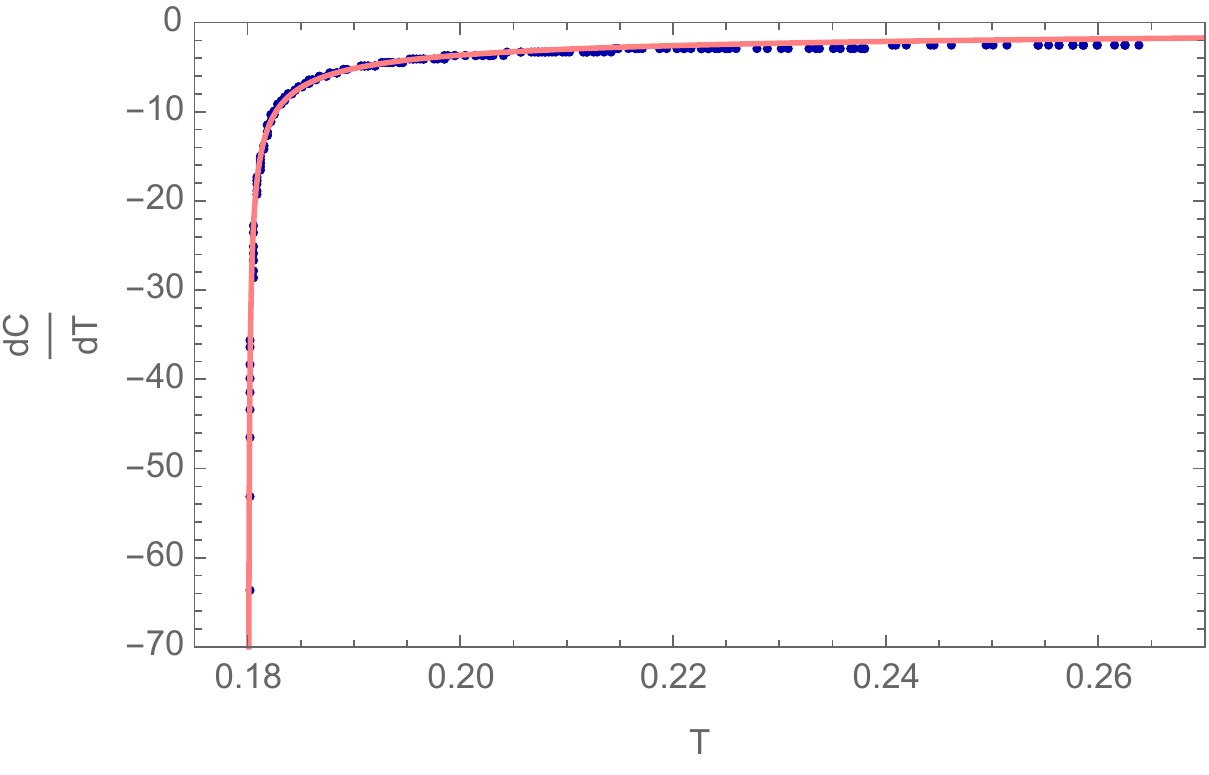}
\includegraphics[width=55 mm]{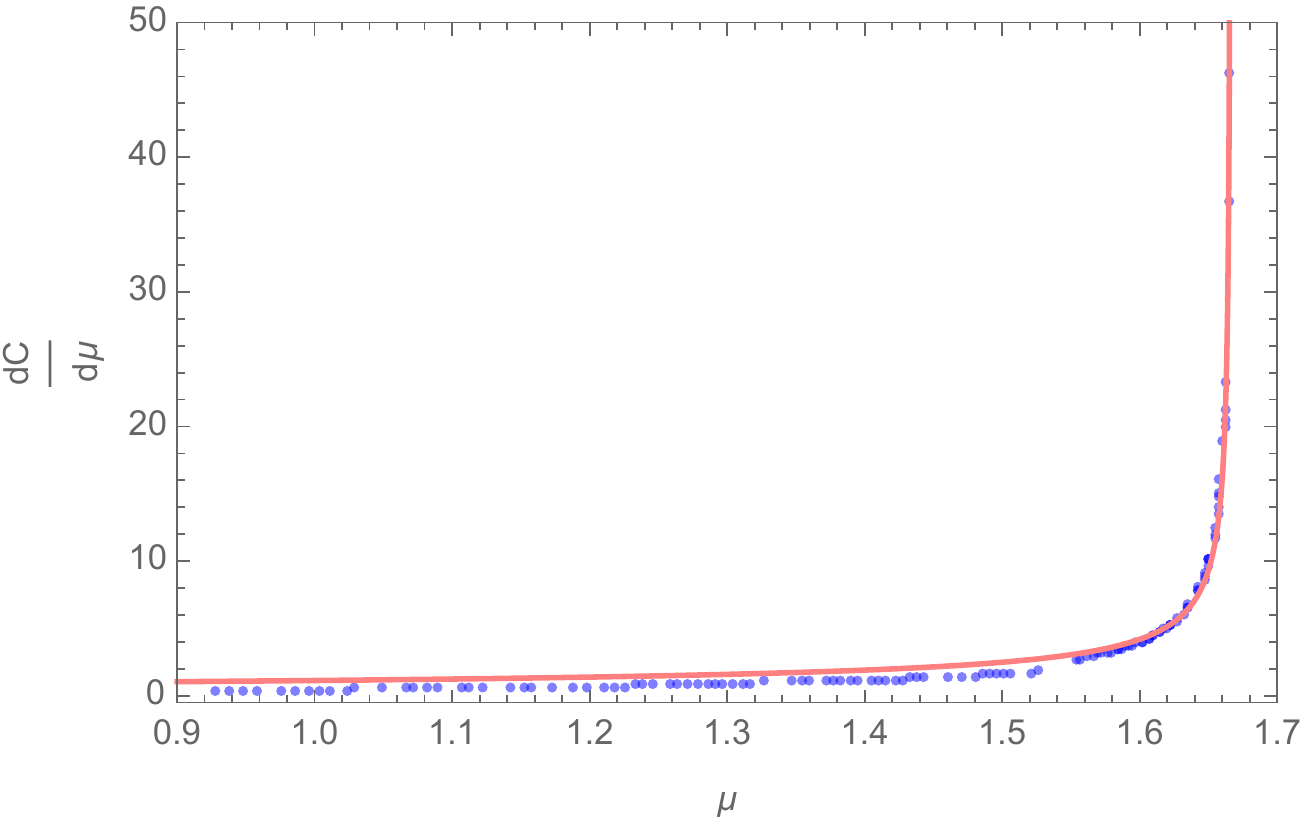}
\includegraphics[width=55 mm]{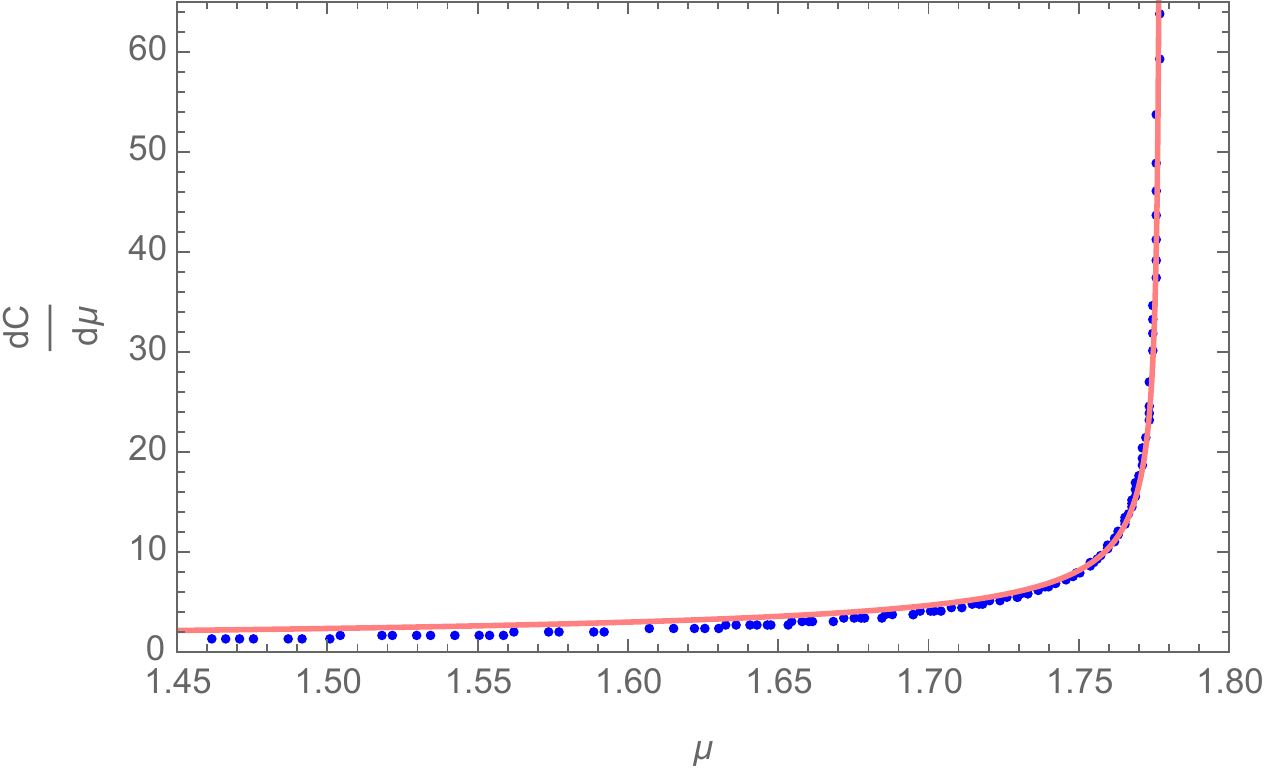}
\includegraphics[width=55 mm]{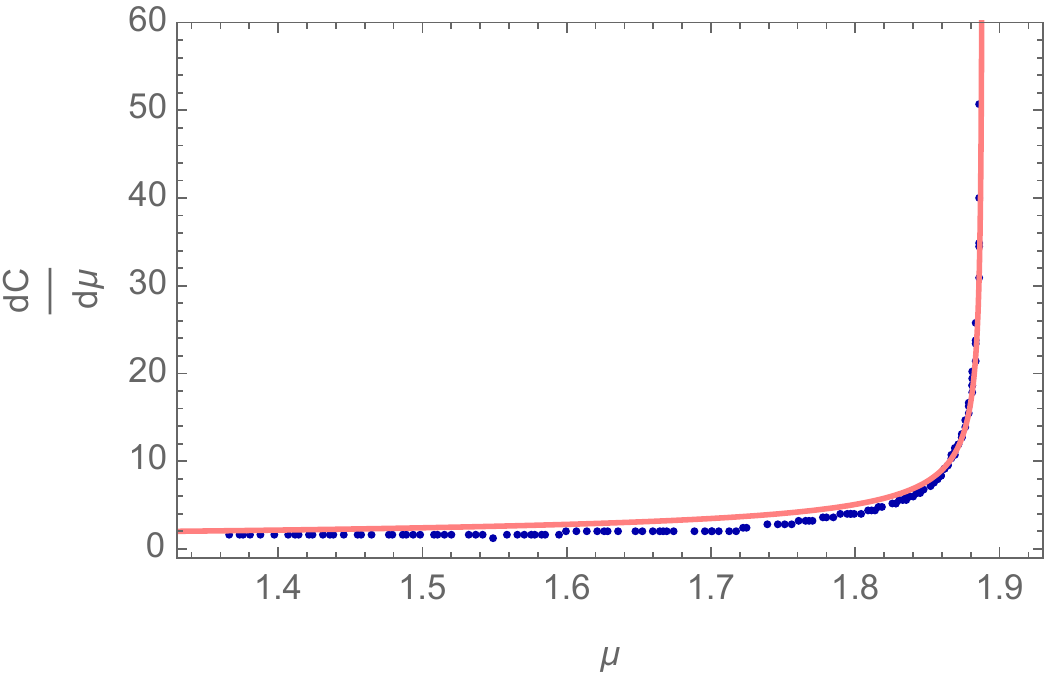}
\caption{The slope of $C$ as a function of $T$ (top row) and $\mu$ (bottom row) for $l=0.1$, near the critical point. The pink curves are the function, $\left(T-T_{c}\right)^{-\theta}$ (top row) and $\left(\mu_{c}-\mu\right)^{-\theta}$ (bottom row), fitted with the data.
Top row: $\mu=0.2$, $T_{c}=0.090032$ and $\theta=0.513351$ (left), $\mu=0.3$, $T_{c}=0.135047$ and $\theta=0.501378$ (middle), $\mu=0.4$, $T_{c}=0.180063$ and $\theta=0.494989$ (right). Bottom row: $T=0.75$, $\mu_{c}=1.666081$ and $\theta=0.566103$ (left), $T=0.8$, $\mu_{c}=1.777153$ and $\theta=0.536357$ (middle), $T=0.85$, $\mu_{c}=1.888225$ and $\theta=0.50306$ (right).}
\label{3}
\end{figure}

As mentioned already, finding critical exponents related to the power laws, which are universal, is one of the most important subjects in the study of critical phenomena. This universality means the values of critical exponents can be independent of the behavior of systems with quite different microscopic properties. Since $C$ diverges near the critical points, as you can see in the low temperature region in the left panel and in the large chemical potential region in the right panel of Fig \ref{2}, an interesting question is whether HSC can give us the critical exponent and consequently information about the phase structure of the theory. We hence define
\begin{align}
\frac{dC}{dT}(i)&=\frac{C(i+1)-C(i)}{T(i+1)-T(i)},\nonumber\\
\frac{dC}{d\mu}(i)&=\frac{C(i+1)-C(i)}{\mu(i+1)-\mu(i)},
\label{slop1}
\end{align}
where $i$ shows the $i$th point of data points. The slope of the curve near the critical points can be fitted with a function of the form $\left(T-T_{c}\right)^{-\theta}$ for the left panel of Fig \ref{2} and $\left(\mu_{c}-\mu\right)^{-\theta}$ for the right panel of Fig \ref{2}. In Fig \ref{3}, we have plotted the slopes \eqref{slop1}, near the critical point. The blue points are numerical results and pink curves are the fitted functions. Our result indicates that although the quantity is quite different from the ones mentioned in papers in the literature, the value of the critical exponent is around 0.5.

\begin{figure}[H]
\centering
\includegraphics[width=67 mm]{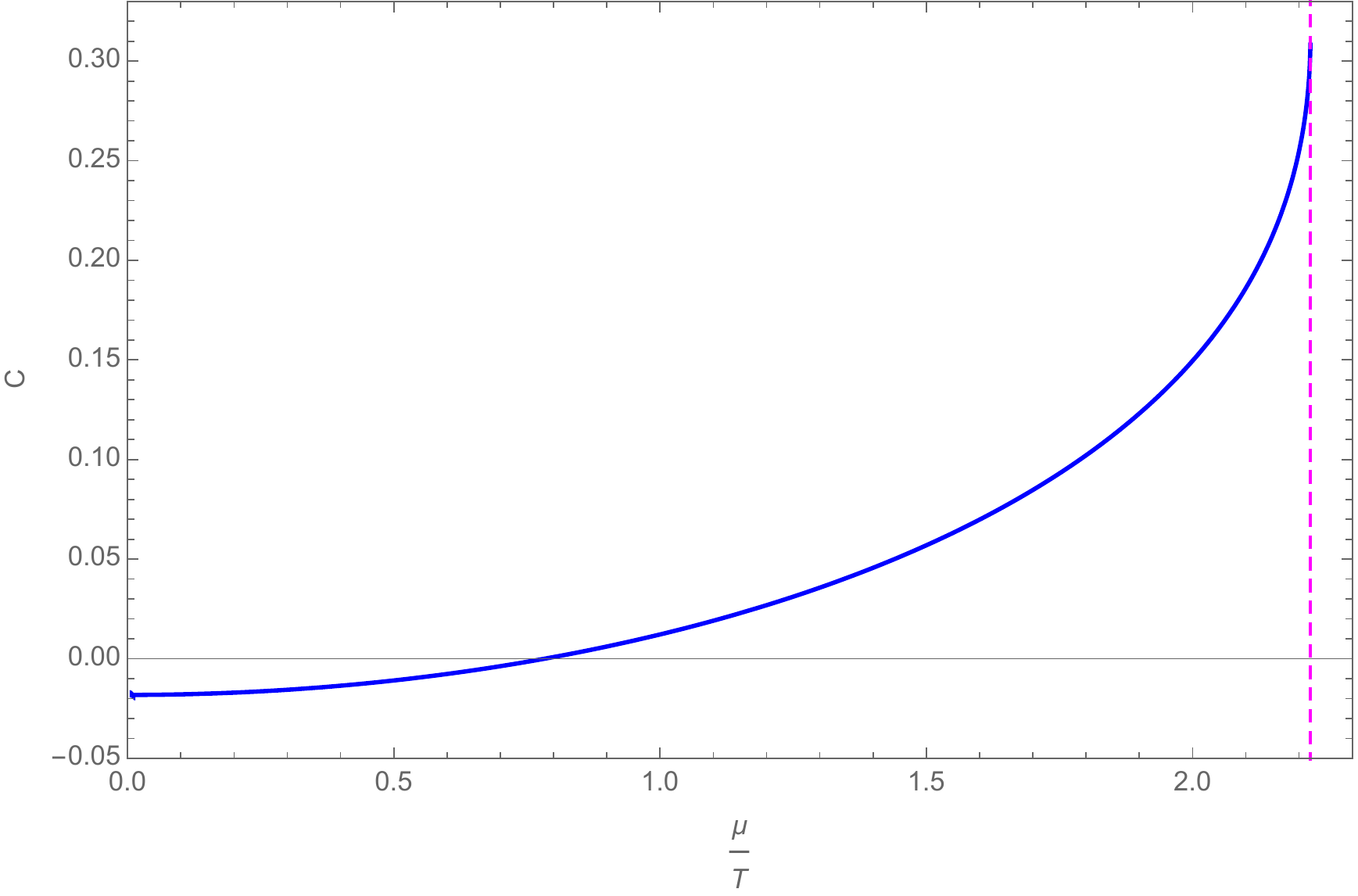} 
\includegraphics[width=67 mm]{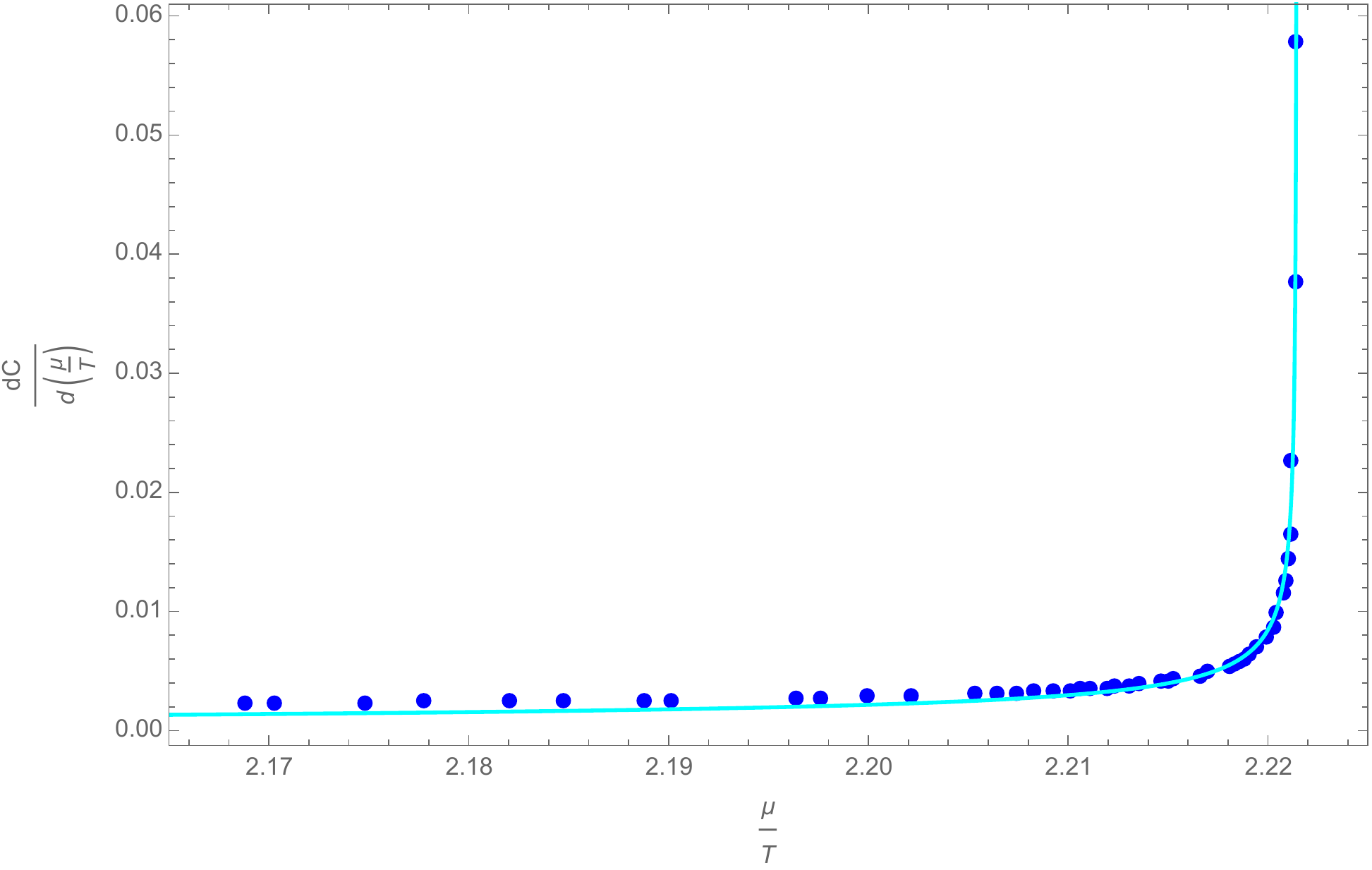} 
\caption{Left: The behavior of $C$ in terms of $\frac{\mu}{T}$ for $l=0.94$. The magenta dashed line shows the critical point which is at $\left(\frac{\mu}{T}\right)_{c}$.  Right: The slope of the value of $C$ as a function of $\frac{\mu}{T}$ for $l=0.94$, near the critical point. The light blue curve is the function, $\left(\frac{\pi}{\sqrt{2}}-\frac{\mu}{T}\right)^{-\theta}$, fitted with the data by $\theta=0.505416$.}
\label{4}
\end{figure}

\begin{figure}[H]
\centering
\includegraphics[width=57 mm]{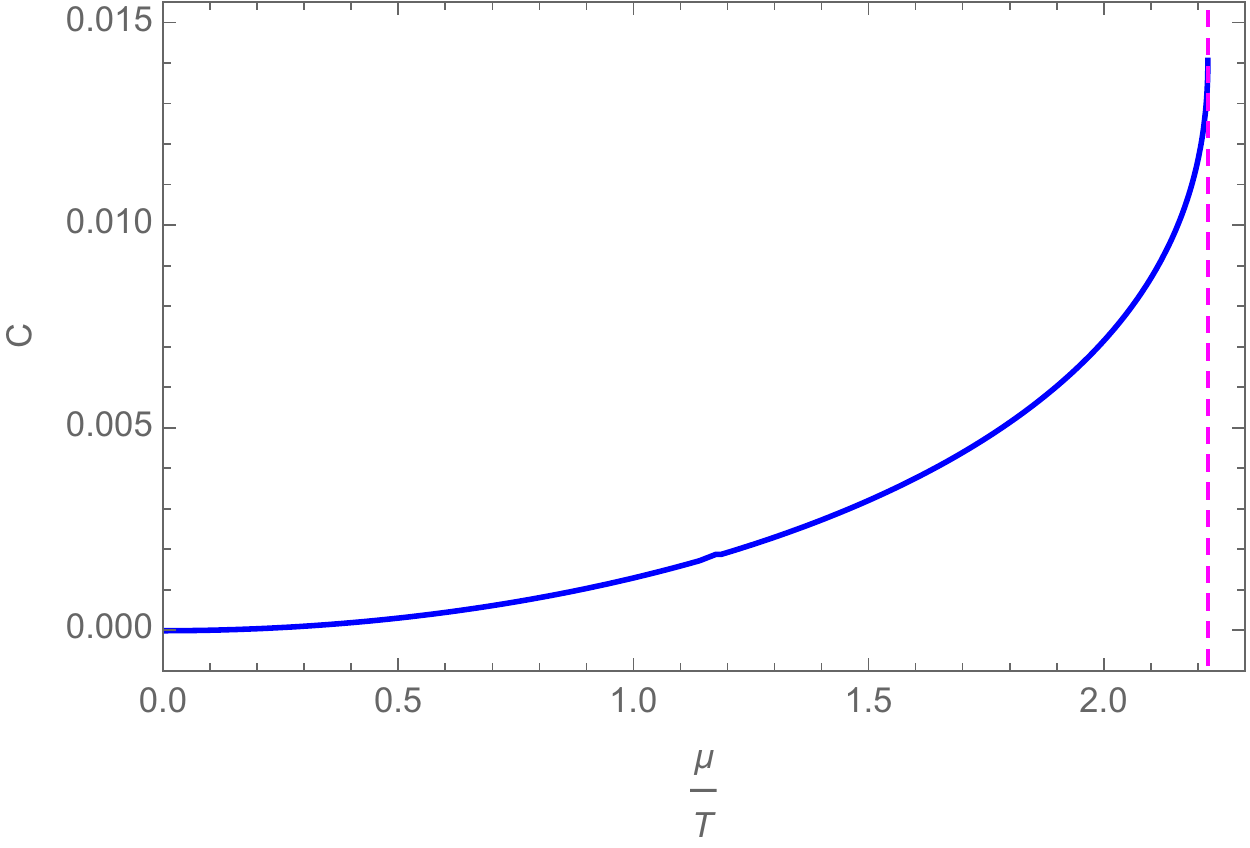} 
\includegraphics[width=57 mm]{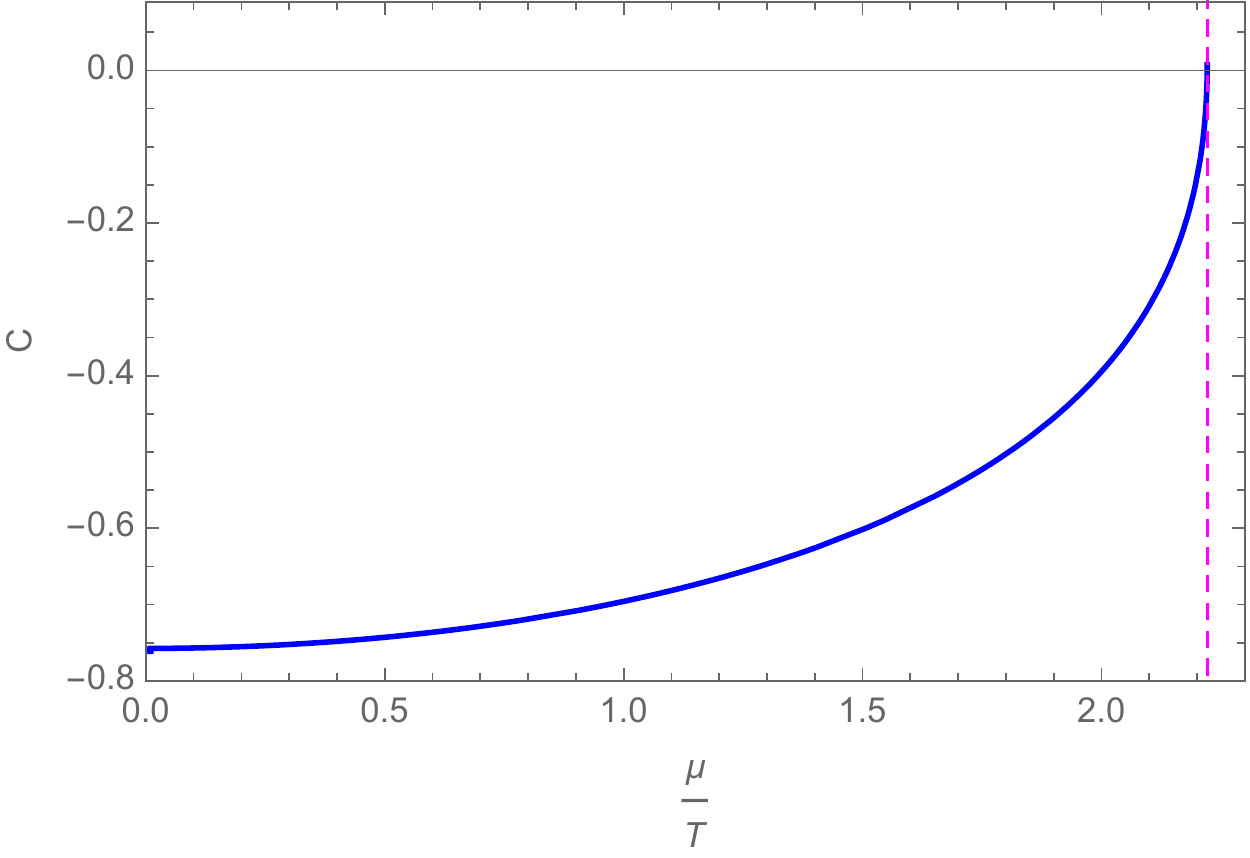} 
\includegraphics[width=57 mm]{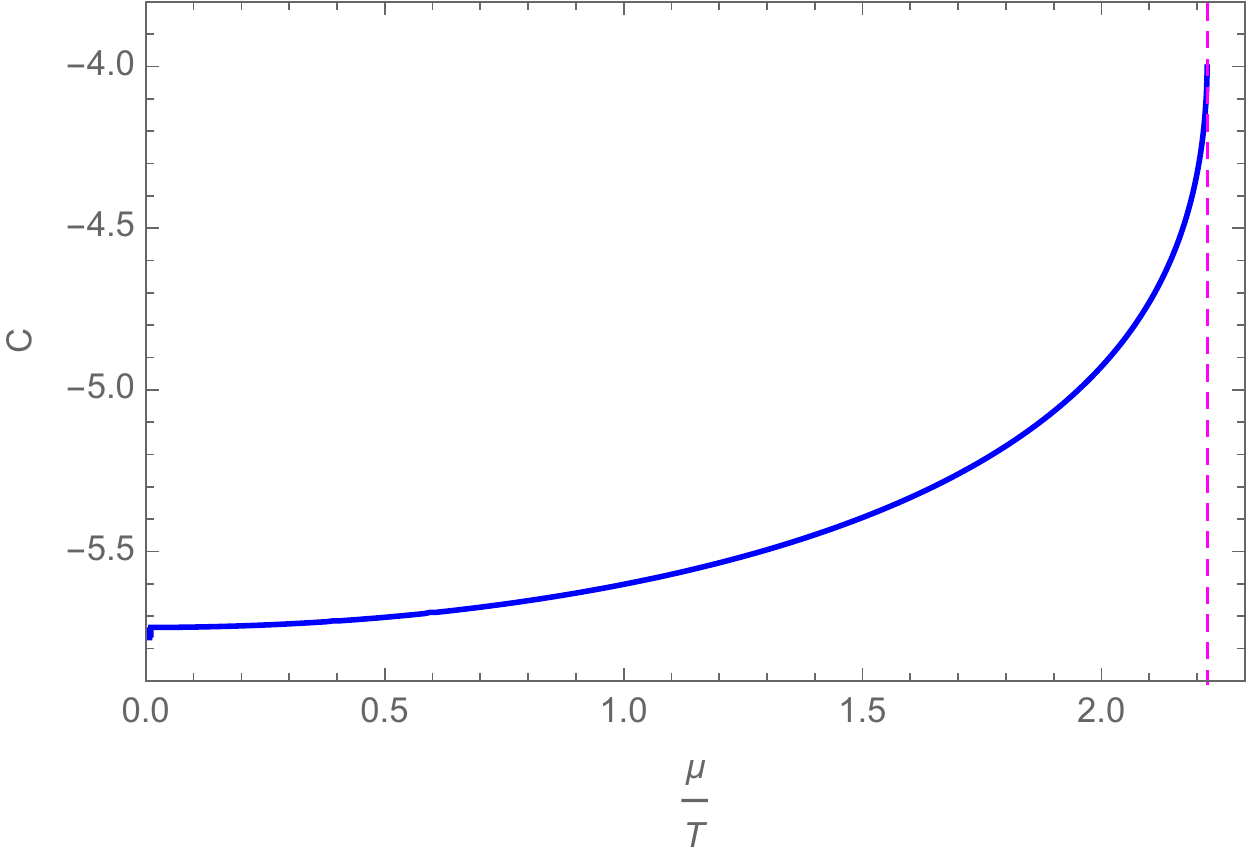} 
\caption{The behavior of $C$ as a function of $\frac{\mu}{T}$ for three different values of $l$ and the critical point marked in the magenta dashed line. Left: $l=0.07$, middle: $l=1.66$ and right: $l=2.41$}
\label{55}
\end{figure}

Since the underlying theory is conformal field theory, reasonably $C$ in analytic calculation must be a function of $\frac{\mu}{T}$. Therefore, we are interested in studying the behavior of $C$  as a function of $\frac{\mu}{T}$. To this end, we have plotted $C$ in terms of $\frac{\mu}{T}$ in the left panel of Fig \ref{4}. 
A quick observation of this figure shows that we have two regions similar to Fig \ref{2}, the $T$ dominated region and the $\mu$ dominated region,
in which $C$ is negative and positive, respectively. Therefore, there is a point in which $C=0$ that means the HSC of two states is the same, according to \eqref{volume2}. We call this point $\left(\frac{\mu}{T}\right)_{*}$. Moreover, when we move towards the critical point in the parameter $\frac{\mu}{T}$, $C$ increases and reaches a finite value with an infinite slope.
Similar to \eqref{slop1}, we can introduce
\begin{align}
&\frac{dC}{d\left(\frac{\mu}{T}\right)}(i)=\frac{C(i+1)-C(i)}{\frac{\mu}{T}(i+1)-\frac{\mu}{T}(i)}.
\label{slope2}
\end{align}

We want to check how the slope changes as a function of $\left(\left(\frac{\mu}{T}\right)_{c}-\frac{\mu}{T}\right)$ this time. The result has been plotted in the right panel of Fig \ref{4} using \eqref{slope2}. The blue points show the numerical result and the light blue curve is the fitted function $\left(\frac{\pi}{\sqrt{2}}-\frac{\mu}{T}\right)^{-\theta}$ with $\theta$ around 0.5. The values of $\theta$ are very close in different ways of calculations, through the dependence of HSC on $T$ and $\mu$ separately and by its dependence on $\frac{\mu}{T}$. 

We want to consider the length of subregion, $l$, as a property of our probe and study the effect of $l$ on exploring of $\left(\frac{\mu}{T}\right)_{*}$, the particular point in which $C=0$. For $l\rightarrow 0$ limit, or $E_{l}(\equiv l^{-1})\rightarrow \infty$, we have $\left(\frac{\mu}{T}\right)_{*}\rightarrow 0$, i.e. there is not a point in which $C$ changes its sign and therefore it is always positive (the left panel of Fig \ref{55}). It means, according to \eqref{volume2}, the HSC of the mixed state in the theory under study is larger than the HSC of the same state in the conformal theory for all values of allowed $\frac{\mu}{T}$. This is because in this UV limit, which is probed by very small $l$, we can probe lower temperatures. Therefore, this limit is equivalent to the $\mu$ dominated region and according to previous explanations, $C$ is always positive. On the other hand, for large enough values of $l$, which means we are probing high temperatures, we have $\left(\frac{\mu}{T}\right)_{*}\rightarrow \left(\frac{\mu}{T}\right)_{c}$ (the middle panel of Fig \ref{55}). In this limit, $\left(\frac{\mu}{T}\right)_{*}$ takes the maximum value it can get, i.e. value of $\frac{\mu}{T}$ in critical point. For IR limit, $l\rightarrow \infty$, or equivalently $E_{l}\rightarrow 0$, there is not any point for changing the sign of $C$ again. But in this limit $C$ is always negative (the right panel of Fig \ref{55}). This is because in this limit, which is probed by very large values of $l$, we can explore very high temperatures. Therefore, this limit is equivalent to the $T$ dominated region from point of view of our probe. You can see a summary of these results in table \eqref{list}.

\begin{table}[ht]
\caption{A short summary of effect of subregion length on $\left(\frac{\mu}{T}\right)_{*}$.}
\vspace{1 mm}
\centering
\begin{tabular}{c c c c}
\hline\hline
~~$ \rm{Sign~of~C} $ ~~   &~~$l (E_{l})$ ~~   &   ~~ $\left(\frac{\mu}{T}\right)_{*}$ ~~  \\[0.5ex]
\hline
$+$ & $ l\rightarrow 0$ $(E_{l}\rightarrow \infty)$ & $\left(\frac{\mu}{T}\right)_{*}\rightarrow 0$  \\
$-$ & $l\gg 1$ $(E_{l}\ll 1)$ & $\left(\frac{\mu}{T}\right)_{*}\rightarrow \left(\frac{\mu}{T}\right)_{c} $ \\
$-$ & $l \rightarrow \infty$ $(E_{l}\rightarrow 0)$ & does not exist   \\
\hline
\end{tabular}\\[1ex]
\label{list}
\end{table}

\begin{figure}[H]
\centering
\includegraphics[width=67 mm]{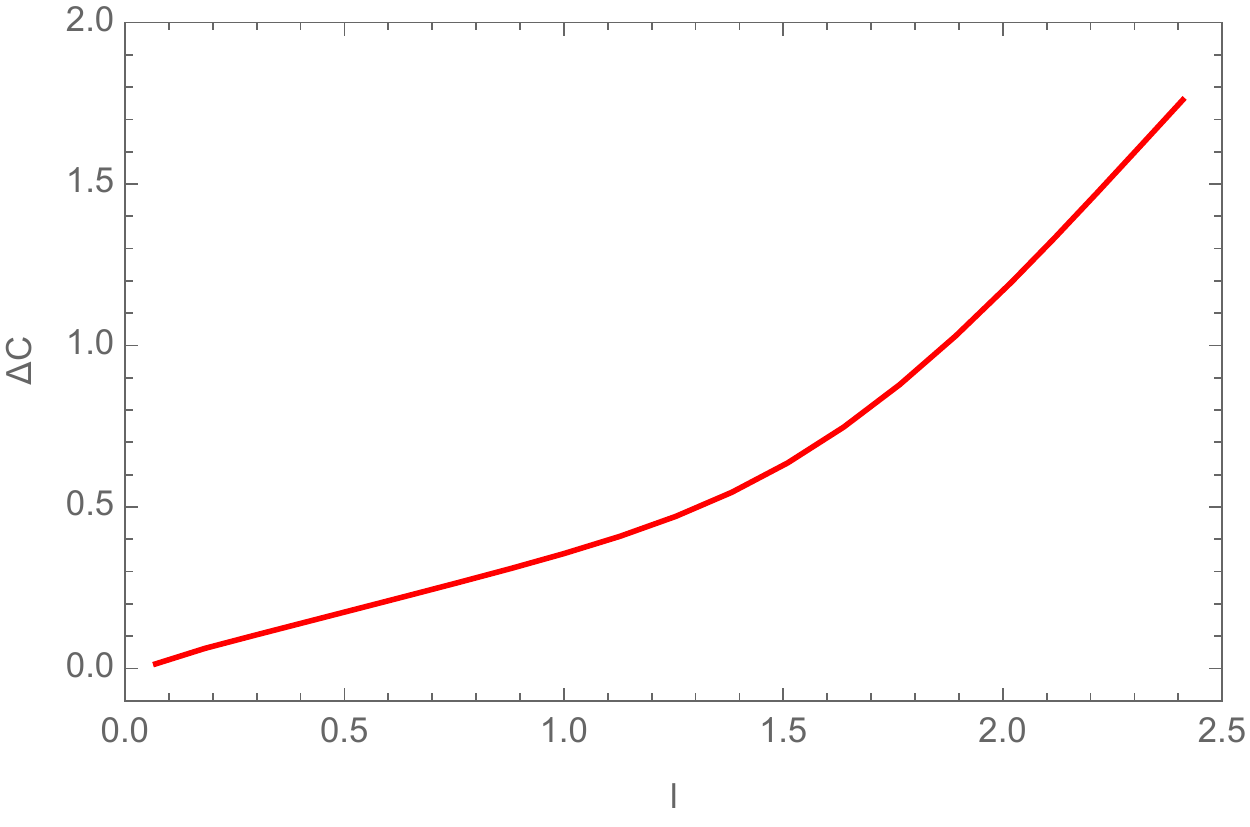} 
\caption{$\Delta C$ as a function of $l$.}
\label{15}
\end{figure}

As mentioned in our introduction, in an analogy with thermodynamic free energy that is used to do a work, uncomplexity, the difference between maximum possible complexity and the actual complexity of a state, is a resource for doing a directed computation \cite{sus1}. In our case, if we consider that the actual complexity is the minimum value of HSC, i.e. the value of HSC in the limit $\frac{\mu}{T}\rightarrow 0$, we can say the computational machine has the maximum resource. In fact, we interpret relative HSC of the initial state, labeled by $\frac{\mu}{T}\rightarrow 0$, and the final state, labeled by $\frac{\mu}{T}\rightarrow \left(\frac{\mu}{T}\right)_{c}$ which has the maximum possible HSC, as a resource which is expended for doing some computational work by a computational machine. Computational work is some computation with a goal. In our definition this goal is to give the initial state to the computational machine and obtain the final state as the output. Therefore,  for a given $l$ we define,

\begin{align}
\Delta C\equiv C|_{\frac{\mu}{T}\rightarrow \left(\frac{\mu}{T}\right)_{c}}-C|_{\frac{\mu}{T}\rightarrow 0}.
\label{difine}
\end{align}
We have plotted the resource, $\Delta C$, as a function of $l$ in Fig \ref{15}. As this figure indicates, by increasing $l$ the needed resource increases, which means we have a better computational machine.

\begin{figure}[H]
\centering
\includegraphics[width=67 mm]{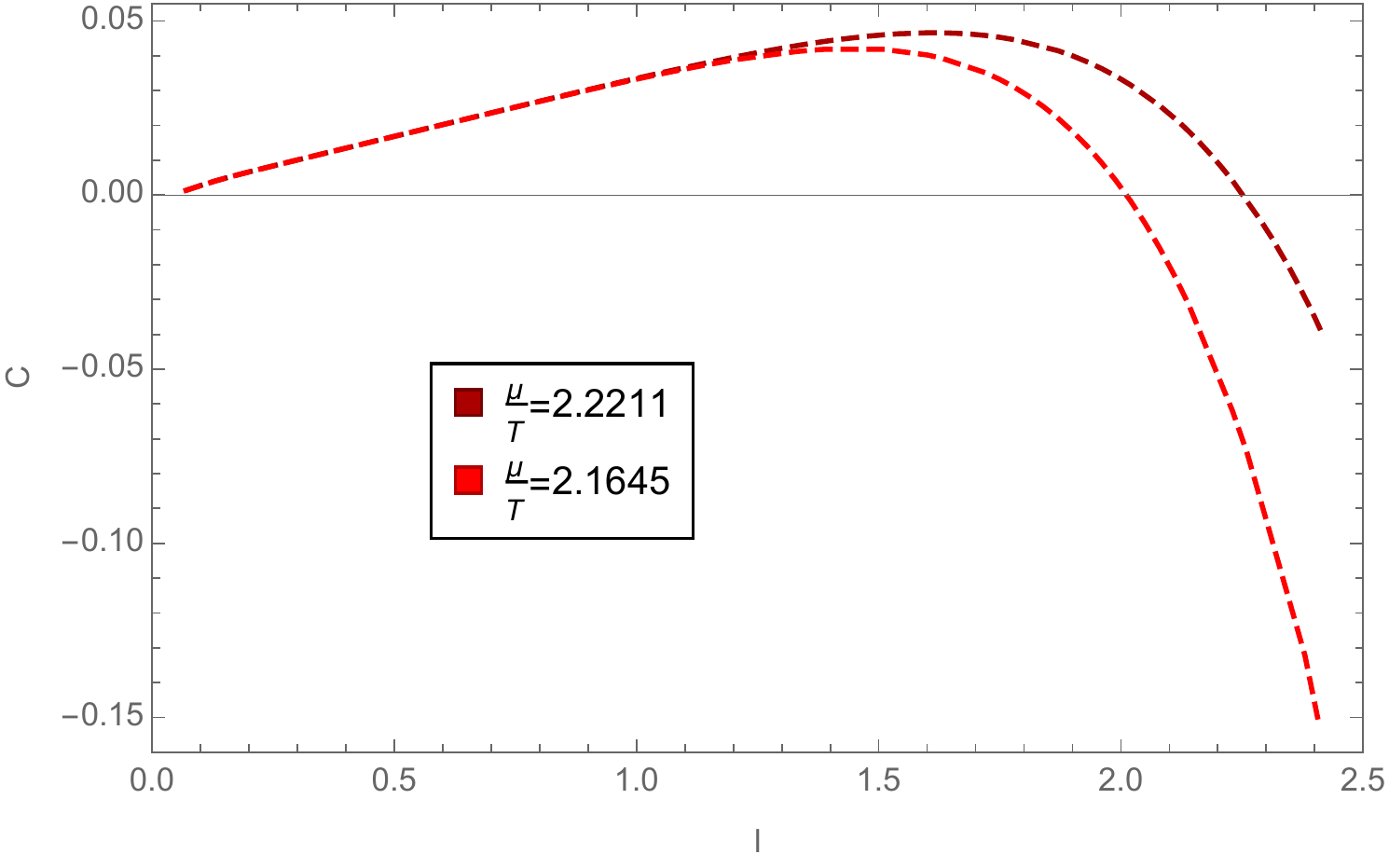} 
\includegraphics[width=67 mm]{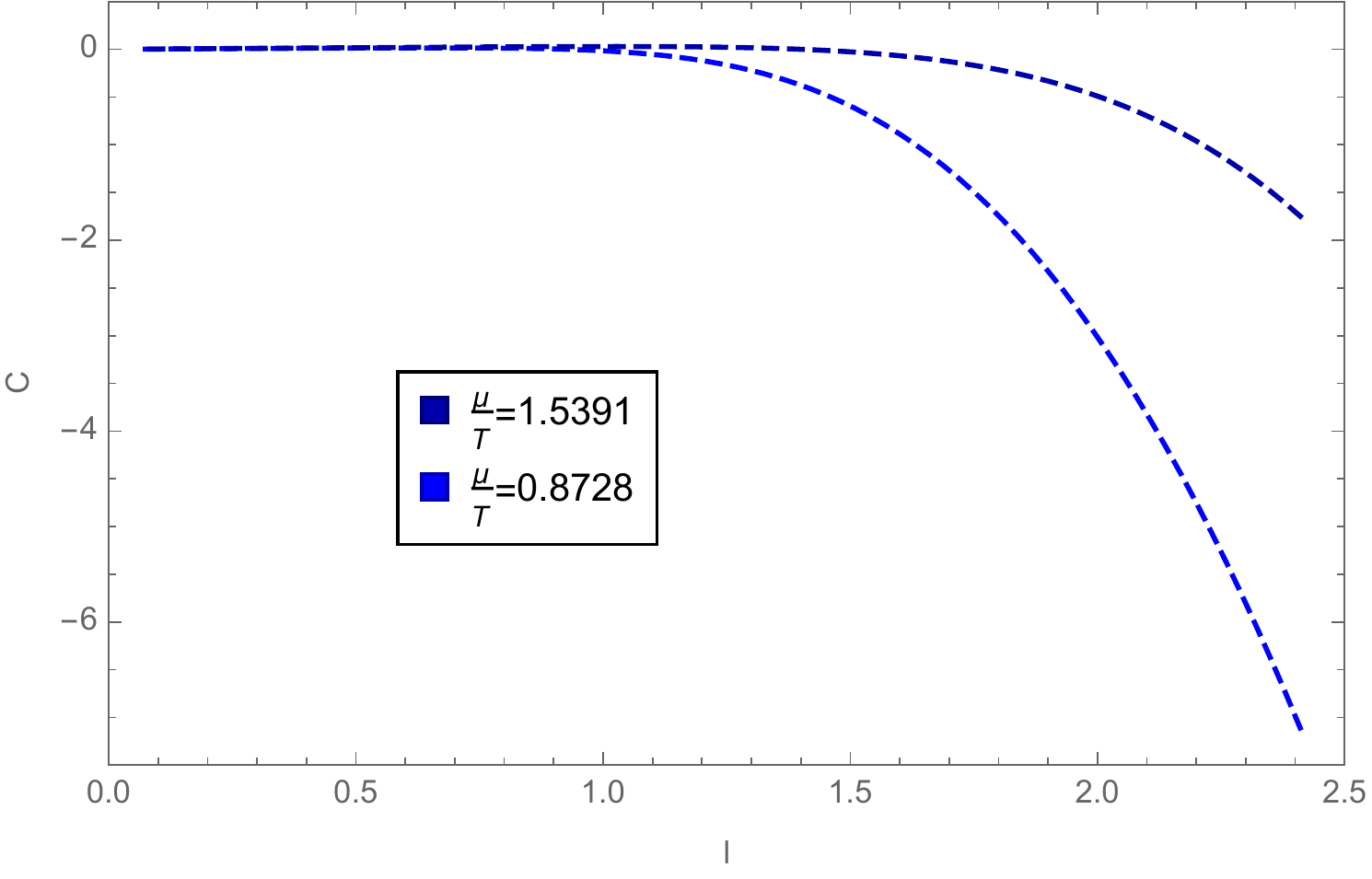} 
\caption{Left: $C$ in terms of $l$ for two fixed values of $\frac{\mu}{T}$ near the critical point. Right: $C$ in terms of $l$ for two fixed values of $\frac{\mu}{T}$ far from the critical point.}
\label{5}
\end{figure}

In Fig \ref{5}, we have plotted $C$ as a function of $l$. This has been shown for two fixed values of $\frac{\mu}{T}$ near the critical point, in the left panel and far from the critical point, in the right panel. As usual, in the UV limit, which is probed by very small $l$, the non-conformal field theory that we study becomes conformal. Therefore, $C$ must go to zero in this limit. In other words, on the gravity side, $r_{*}$ is close to the boundary when $l$ has very small values. Near the boundary region is $AdS_{5}$ with good accuracy which contributes more to $\mathcal{C}$. As $l$ increases, $r_{*}$ reaches deeper in the bulk and nears the horizon, so the deviation of geometry from $AdS_{5}$ can be probed further by minimal surface or equivalently HSC. As you can see in the left panel, in the small enough values of $l$, when $\mu$ is about twice $T$ i.e. near the critical point, the effect of $\mu$ prevails and $C$ is positive. With the increase of $l$, thermal effects become dominant and $C$ becomes negative. However, by moving away from the critical point, at smaller values of $\frac{\mu}{T}$, even in a small region of $l$, $C$ does not show a change in sign. Moreover, as this figure shows, at fixed $l$, by comparing non-conformal states, the closer we are to the critical point, the more information is needed to specify the state.

\section*{Acknowledgements}
This work is based upon research funded by Iran National Science Foundation (INSF) under Project No. 98013297.

\end{document}